\documentclass[floatfix,aps,prapplied,preprint, superscriptaddress]{revtex4-2}

\usepackage[pdftex]{graphicx}
\usepackage{times,amsmath,amssymb}
\usepackage{color}
\usepackage[colorlinks = true,
            linkcolor = blue,
            urlcolor  = blue,
            citecolor = blue,
            anchorcolor = blue]{hyperref}

\begin{document}

\title{Charge state estimation in quantum dots using a Bayesian approach}

\author{Motoya Shinozaki}
\email[]{motoya.shinozaki.c1@tohoku.ac.jp}
\affiliation{WPI-Advanced Institute for Materials Research, Tohoku University, 2-1-1 Katahira, Aoba-ku, Sendai 980-8577, Japan}

\author{Yui Muto}
\affiliation{Research Institute of Electrical Communication, Tohoku University, 2-1-1 Katahira, Aoba-ku, Sendai 980-8577, Japan}
\affiliation{Graduate School of Engineering, Tohoku University, 6-6 Aramaki Aza Aoba, Aoba-ku, Sendai 980-0845, Japan}

\author{Takahito Kitada}
\affiliation{Research Institute of Electrical Communication, Tohoku University, 2-1-1 Katahira, Aoba-ku, Sendai 980-8577, Japan}
\affiliation{Graduate School of Engineering, Tohoku University, 6-6 Aramaki Aza Aoba, Aoba-ku, Sendai 980-0845, Japan}

\author{Tomohiro Otsuka}
\email[]{tomohiro.otsuka@tohoku.ac.jp}
\affiliation{WPI-Advanced Institute for Materials Research, Tohoku University, 2-1-1 Katahira, Aoba-ku, Sendai 980-8577, Japan}
\affiliation{Research Institute of Electrical Communication, Tohoku University, 2-1-1 Katahira, Aoba-ku, Sendai 980-8577, Japan}
\affiliation{Department of Electronic Engineering, Graduate School of Engineering, Tohoku University, Aoba 6-6-05, Aramaki, Aoba-Ku, Sendai 980-8579, Japan}
\affiliation{Center for Science and Innovation in Spintronics, Tohoku University, 2-1-1 Katahira, Aoba-ku, Sendai 980-8577, Japan}
\affiliation{Center for Emergent Matter Science, RIKEN, 2-1 Hirosawa, Wako, Saitama 351-0198, Japan}

\date{\today}

\begin{abstract}
Detection of single-electron charges in solid-state nanodevices is a key technique in semiconductor quantum bit readout for quantum information processing and probing electronic properties of nanostructures.
This detection is achieved using quantum dot charge sensors, with its speed enhanced by high-speed RF reflectometry. 
Recently, real-time processing of data from RF reflectometry has attracted much attention to quantum information processing.
In this paper, we propose a sequential method based on Bayes' theorem for estimating the charge state and compare its performance with the averaging approach and threshold judgment.
When the noise variance differs between the empty and occupied states, the Bayesian approach demonstrates a lower error score, facilitating the extraction of more data points in real-time charge state estimation. 
Additionally, the Bayesian approach outperforms the averaging method and threshold judgment in terms of error rates for charge state estimation, even during charge transitions. 
This technique is broadly applicable to single-electron detection and holds substantial utility for quantum bit readout and the operation of nanoprobes that utilize single-electron detection.
\end{abstract}

\maketitle

\section{Introduction}

The detection of single-charge transitions in solid-state nanodevices holds significant importance in both basic science and device applications.
The measurement of electron movement within nanostructures unveils the electronic properties of these structures, such as quantum states formed inside them~\cite{1996TaruchaPRL, 1997KouwenhovenSci, 2000CiorgaPRB, 2010AltimirasNatPhys, 2012OtsukaPRB}.
Moreover, it facilitates the exploration of intriguing physical phenomena like the full counting statistics of current~\cite{2006FujisawaScience, 2006GustavssonPRL} and demonstrations of Maxwell's demon~\cite{2017ChidaNatCom}. 
Control over single-charge movements in nanostructures also enables practical applications, such as current standards~\cite{2013PekolaRMP, 2017YamahataSciRep}.
In quantum information processing utilizing electron spins in quantum dots (QDs)~\cite{1998LossPRA, 2010LaddNat, 2013AwschalomSci}, the single charge detection is used in the quantum bit (qubit) readout.
This technique allows for the projection of spin qubit information onto electron charge states, as demonstrated in various reports\cite{2005PettaSci, 2006KoppensNat, 2014YonedaPRL, 2018TakedaNpjQ, 2021KojimaNpjQ}.

To detect single-electron charges in nanostructures, charge sensors utilizing quantum point contacts (QPCs) or QDs are powerful tools~\cite{1993FieldPRL, 1998BuksNat, 2002SprinzakPRL, 2003ElzermanPRB}.
It becomes possible to speed up the operation of the sensors and improve the data acquisition rate by utilizing a radio-frequency (RF) technique called RF-reflectometry~\cite{1998SchelkopfSci, 2007ReillyAPL, 2010BarthelPRB}.
This technique is nowadays used for fast readout of electron spin qubits for quantum computing~\cite{2014ShulmanNatCom, 2018YonedaNatureNano, 2019NakajimaNatureNano, yoneda2020quantum, noiri2020radio, johmen2023radio} and fast operation of nanoelectronic probes in solid states~\cite{2015OtsukaSciRep, 2017OtsukaSciRep}.
To estimate the charge state, threshold judgment is commonly employed with an enhancement of the signal-to-noise ratio through time integration~\cite{barthel2009rapid, 2010BarthelPRB, 2015OtsukaSciRep, noiri2020radio, johmen2023radio}. 
This time integration results in limitations in the readout speed.
To address this issue, the development of ultra-low noise amplifiers has become mainstream~\cite{yamamoto2008flux, stehlik2015fast}, as amplifier noise is a dominant factor in RF-reflectometry~\cite{shinozaki2021gate}.

In this paper, we address this issue from the perspective of data analysis algorithms. 
One promising candidate is the Bayesian approach, utilized in the creation of new materials as referenced in various studies~\cite{2015RajanARMR, 2005RajanIMT, 2016AgrawalAPLM, 2016UenoMD, 2017JuPRX}, and also in single-shot spin readout~\cite{struck2020robust, mizokuchi2020detection}.
Furthermore, field-programmable gate arrays (FPGAs) have enabled high-speed and real-time data processing and have been applied to feedback control in quantum dots~\cite{2014ShulmanNatCom, nakajima2020coherence, nakajima2021real, fujiwara2023wide}.
In light of this background, we propose a sequential real-time estimation method based on Bayes' theorem, designed to be compatible with FPGAs for accelerated real-time processing and compare the performance of a sequential processing method utilizing averaging with that of the traditional threshold judgment approach.

\begin{figure*}[t]
\begin{center}
  \includegraphics{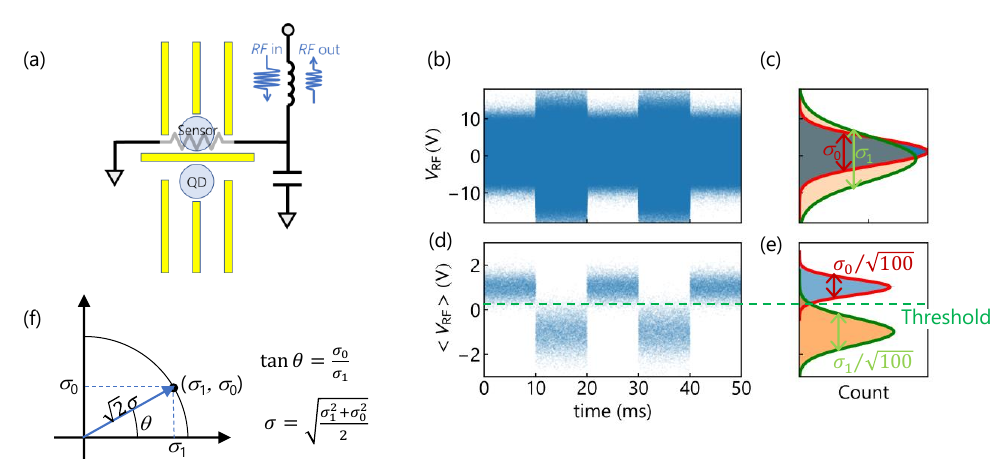}
  \caption{(a) Schematic of the measurement setup considered in a simulation.
  A sensor QD is embedded in an RF tank circuit for the RF reflectometry.
  (b), (c) Example of simulated data of the $V_{\rm RF}$ as a function of the time and histogram of the $V_{\rm RF}$.
  The noise distribution $\sigma $ is larger than the signal difference between the empty state (0) and the occupied state (1).
  (d), (e) Averaged data $\langle V_{\rm RF}\rangle $ as a function of the time and histogram of the $\langle V_{\rm RF}\rangle $.
  Averaging is done by using $N=$100 data points.
  The effective noise distribution becomes small.
  (f) A relationship of two noise distributions in the simulation.
  }
  \label{Signal}
\end{center}
\end{figure*}
\section{Results}
\subsection{Simulation setup}

Figure~\ref{Signal}(a) shows a schematic of the QD charge sensor.
A sensor QD (upper) is placed next to a target QD (lower).
Charging of the target QD by a single electron modifies the electrostatic potential at the sensor and its conductance~\cite{2010BarthelPRB}.
The sensor conductance when the target QD is empty (0) is different from the conductance when the QD is occupied (1).
The sensor QD is embedded in an RF tank circuit.
The resonance condition of the circuit is modified by the change of the sensor conductance and then the reflected RF voltage $V_{\rm RF}$ is modified.
We simulate the $V_{\rm RF}$ in every 8 ns corresponding to a sampling rate of 125MHz.

Figure~\ref{Signal}(b) and (c) show the simulated $V_{\rm RF}$ as a function of the time and a histogram of $V_{\rm RF}$.
We assume a white Gaussian noise in the simulation here to simplify the discussion.
We discuss cases with frequency-dependent noise~\cite{shinozaki2021gate} in Appendix A.
The noise distribution $\sigma $ is larger than the signal separation between the empty state (0) and the occupied state (1) $\Delta V_{\rm RF}=|V_{\rm RF1}-V_{\rm RF0}|$.
Also, we assume the time scale of the charge state transition is much slower than the measurement repetition 8~ns.
This condition is realized by choosing the slow dynamics condition with small tunnel coupling to the leads in real measurement.
The noise for 0 and 1 states ($\sigma_0$ and $\sigma_1$) will be affected by the charge sensitivity of the sensor at 0 and 1 states \cite{eenink2019tunable, shinozaki2021gate}. 
In a highly sensitive charge sensor, $\sigma_0$ and $\sigma_1$ can be different values because of the shift of the operation point.

The simplest method to extract the charge state from this noisy data would be averaging and threshold judgment.
By calculating arithmetic means using $N$ data points, the effective noise distribution shrinks to $\sigma/\sqrt{N}$.
Figures~\ref{Signal}(d) and (e) show the averaged RF signal $\langle V_{\rm RF} \rangle $ and a histogram of $\langle V_{\rm RF} \rangle $ with $N=100$.
Now, the noise distribution becomes smaller than the signal separation, and we can see the transition of the charge state between 0 and 1 by setting a threshold at the middle point between the two peaks (Fig.~\ref{Signal}(e)).
This method will be discussed in further detail later in subsection C.

\begin{figure*}[t]
\begin{center}
  \includegraphics{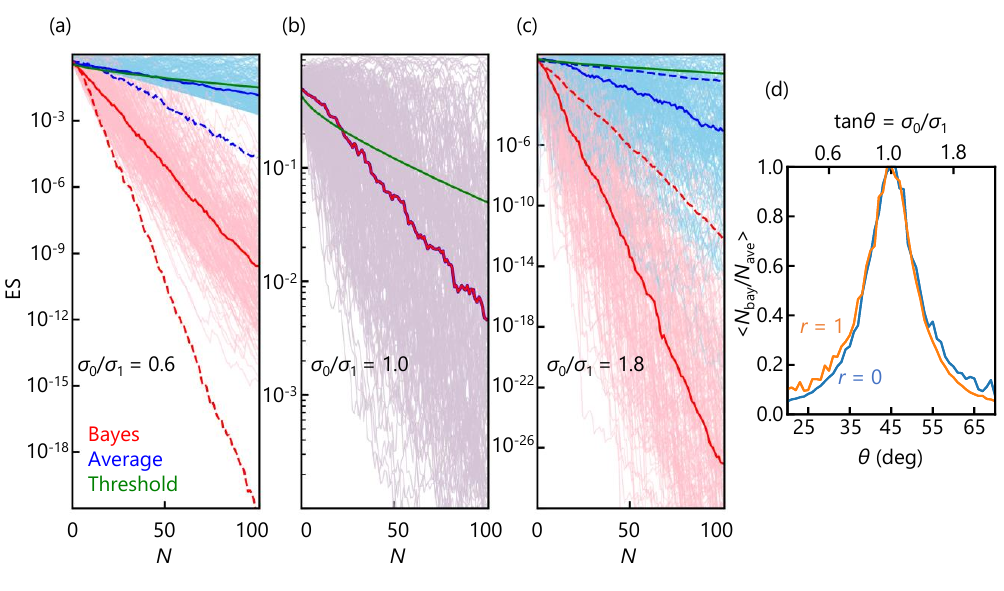}
  \caption{(a)-(c) Simulated error score (ES) as a function of the number of data points used for the charge state estimation $N$.
The pink traces show $\rm{ES}_{0,\rm bay}$.
The traces show the results with 200 different datasets.
The red trace shows the median of the datasets.
The light blue traces show $\rm{ES}_{0,\rm ave}$ calculated by Eq.~\ref{eq:Ave}
The blue trace shows the median of the datasets.
  The signal-to-noise ratio ($\rm{SNR}=\Delta V_{\rm RF}/\sigma$) is fixed to 0.33.
  The dashed lines indicate the $\rm{ES}_{1}$.
  The noise distribution ratio is assumed as $\sigma_0/\sigma_1 = 0.6, 1, 1.8$ in (a), (b) and (c), respectively.
  (d) The noise ratio dependence of sampling numbers ratio $N_{\rm bay}/N_{\rm ave}$ for archiving the $\rm{ES}_{0}$ of $10^{-4}$. The small $N_{\rm bay}/N_{\rm ave}$ indicates that the Bayesian approach has better performance than the averaging approach. The blue and orange lines correspond to the estimated states 0 and 1.
  }
  \label{Simulation}
\end{center}
\end{figure*}
\subsection{Bayesian approach}
In the charge state estimation, we can improve the performance by utilizing the Bayesian approach.
We possess the entire dataset as a function of time, denoted as $\mathbf{V_{\rm RF}}= \{ V_{{\rm RF\_} 0}, \ldots, V_{{\rm RF\_} n}\}$, along with information on the distributions $\sigma_0, \sigma_1$ and the signals $V_{\rm RF0}, V_{\rm RF1}$ obtained from a calibration measurement.
By utilizing Bayes' theorem, the probability that the charge state is 0 given the obtained data $\mathbf{V_{\rm RF}}$ becomes
\begin{equation}
P(0|\mathbf{V_{\rm RF}}) = \frac{P(\mathbf{V_{\rm RF}}|0)P(0)}{P(\mathbf{V_{\rm RF}})}.
\label{eq:Bayes}
\end{equation}
Here, $P(\mathbf{V_{\rm RF}}|0)$ is the probability to obtain $\mathbf{V_{\rm RF}}$ when the state is 0, $P(0)$ is the probability that the state is 0 and $P(\mathbf{V_{\rm RF}})$ is the probability to obtain $\mathbf{V_{\rm RF}}$.
We can calculate this probability by using the dataset with $P(\mathbf{V_{\rm RF}}|0)=\prod_{i=0}^n P(V_{\rm RF\_i}|0)$ and $P(\mathbf{V_{\rm RF}})=P(0)\prod_{i=0}^n P(V_{\rm RF\_i}|0)+P(1)\prod_{i=0}^n P(V_{\rm RF\_i}|1)$ if the noise is white Gaussian.

We utilize this probability in the charge state estimation.
When we denote the actual charge state by $q$, and the estimated state $r$ is 1, the error score that $q$ is not equal to $r$ is becomes
\begin{equation}
{\rm ES}_{0, \rm bay}=P(r=1|\mathbf{V_{\rm RF}}) = 1- P(r=0|\mathbf{V_{\rm RF}}).
\end{equation}
By increasing the number of data points $n$, we can calculate the error score sequentially.
When ${\rm ES}_{0, \rm bay}$ becomes smaller than the target value that we set, we finish the estimation as the estimated state is 0.
(On the other hand, ${\rm ES}_{1, \rm bay}$ becomes smaller than the target value, and we set the estimated state as 1.)
Note that the error score is generally different from the error rate $P(r\neq q|q)$.
Later, we will discuss the relation and how we can realize the state estimation satisfying specific error rates by monitoring the error score.

We compare the performance with the sequential averaging approach in which we utilize averaging to calculate the error score.
By considering the reduction of $\sigma$ in Fig.~\ref{Signal}(e), the error score in the averaging approach for 0 state becomes
\begin{equation}
{\rm ES}_{0,\rm ave}=1-\frac{P_{\rm N}(0)}{P_{\rm N}(0)+P_{\rm N}(1)} = \frac{1}{1+\frac{P_{\rm N}(0)}{P_{\rm N}(1)}}
\label{eq:Ave}
\end{equation}
where, $P_{\rm N}(q)$ is the normal distribution $(\frac{1}{\sqrt{2\pi}\sigma} \rm exp \left\{-\frac{\left(\langle V_{\rm{RF}}\rangle-V_{{\rm RF}q}\right)^2}{2 \sigma^2}\right\}$ with the distribution $\sigma = \sigma_q /\sqrt{N}$.
We have assumed the noise is the white Gaussian.

\begin{figure}
\begin{center}
  \includegraphics{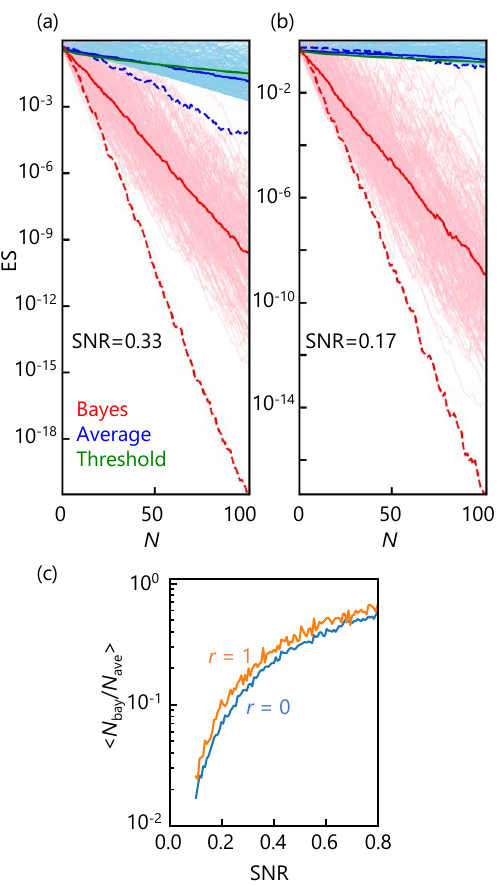}
  \caption{Calculated $\rm{ES}_{0}$ in the case of $\rm{SNR}$ = 0.33 (a) and 0.17 (b), respectively. The dashed line indicates the $\rm{ES}_{1}$. (c) The $\rm{SNR}$ dependence of $N_{\rm bay}/N_{\rm ave}$.
  $N_{\rm bay}/N_{\rm ave}$ decreases with the decrease of $\rm{SNR}$.
  }
  \label{Experiment}
\end{center}
\end{figure}
Figure~\ref{Simulation} (a-c) shows calculated ES as a function of the number of data points used for the charge state estimation $N$.
Here the data for this estimation is created by adding a normal distribution noise with the distributions $\sigma_{0,1} $ and signal-to-noise ratio ($\rm{SNR}=\Delta V_{\rm RF}/\sigma$) is fixed to 0.33.
The noise distribution ratio is $\sigma_0/\sigma_1 = 0.6, 1.0,$ and $1.8$ in (a), (b), and (c), respectively.
The probability in the estimation $P(q)$ is assumed to be 0.5.

The pink traces show $\rm{ES}_{0,\rm bay}$.
The traces show the results with 200 different datasets.
The red trace shows the median of the datasets.
The light blue traces show $\rm{ES}_{0,\rm ave}$ calculated by Eq.~\ref{eq:Ave}.
The blue trace shows the median of the datasets.
The $\rm ES_{0, ave}$ decreases with increasing $N$ by $1/\sqrt{N}$ because $\sigma_{0, 1}$ shrink by the averaging.
We also show the median of ${\rm ES}_{1}$ for $q=1$ in cases of the Bayesian and averaging approaches as the red and blue dashed lines. 
The median value of ES (when $\sigma_0$ and $\sigma_1$ are not equal) of the Bayesian approach is smaller than that of the averaging approach.
In the case of $\sigma_0 \approx \sigma_1$ (Fig.~\ref{Simulation}(b)), both the Bayesian and the averaging approaches show the same performance.

We summarize the comparison of the Bayesian and the averaging methods in noise ratio dependence in Fig.~\ref{Simulation}(d). 
The $N_{\rm bay}$ and $N_{\rm ave}$ are defined as the numbers of the required sampling data points satisfying the $\rm{ES}_{0}$ of $10^{-4}$.
The Bayesian approach shows better performance with a more unbalanced noise state. 
For example in the case of $\sigma_0/\sigma_1=0.6$, $\langle N_{\rm bay} \rangle$ is about 10 times smaller than $\langle N_{\rm ave} \rangle$ for $r=0$.
In the case of $\sigma_0<\sigma_1$ (such as the condition shown in Fig.~\ref{Simulation}(a)), detecting the value likely attributed to the state 0 is more informative than when reading the measurement of state 1, so the ${\rm ES}_{1}$ is lower than ${\rm ES}_{0}$.
Therefore, the asymmetric behavior in Fig.~\ref{Simulation}(d) is observed.
Note that $N_{\rm bay}$ we discussed here, which only treats one of $\rm ES$s, is a conservative evaluation because $N_{\rm bay}$ becomes smaller with treating both $\rm ES_{0}$ and $\rm ES_{1}$.

Next, we investigate the signal-to-noise ratio ($\rm{SNR}$) dependence. 
The ratio of the noise distribution is fixed to 0.6 in the simulation shown in Fig.~\ref{Experiment}.
Figures~\ref{Experiment} (a) and (b) show the results with $\rm{SNR}$ = 0.33 and 0.17, respectively.
In both cases, $\rm{ES}$ of the Bayesian approach is lower than that of the averaging approach.
The $\rm{ES}_{1}$ are also shown as dashed line in Fig.~\ref{Experiment} (a) and (b).
$\langle N_{\rm bay} \rangle/\langle N_{\rm ave}\rangle$ of $r=0$ as a function of $\rm{SNR}$ is shown in Fig.~\ref{Experiment} (c).
$\langle N_{\rm bay} \rangle /\langle N_{\rm ave}\rangle$ decreases with the decrease of $\rm{SNR}$, which indicates that the Bayesian approach shows better performance in the case of larger noise intensity and state-dependent noise.

\begin{figure}
\begin{center}
  \includegraphics{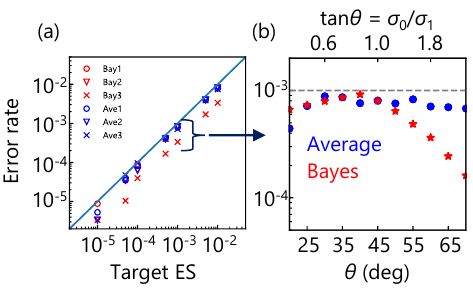}
  \caption{(a) The error rate of Baysian and averaging approach as a function of a target ES. The conditions 1, 2, and 3 correspond to the $\sigma_0/\sigma_1 = 0.6, 1, 1.8$, respectively. 
  (b) A noise ratio dependence of the error rate of the Bayesian and averaging approach.
  The error rates are evaluated for $q=0$.
  }
  \label{error}
\end{center}
\end{figure}

\subsection{Error rate of the Bayesian approach}
Here, we evaluate the error rates of the Bayesian approach and the averaging approach and compare those with threshold judgment.
We count up the number of failure events with estimated states as 1 when we prepare the state as 0 by using datasets of $6.25 \times 10^{7}$ points.
First, we evaluate the error rate as a function of the target value in the state estimation utilizing ES (target ES) as shown in Fig~\ref{error}(a).
We evaluate in three conditions with $\sigma_0/\sigma_1 = 0.6, 1.0$, and $1.8$ labeled as 1, 2, and 3, respectively.
The error rate decreases with the decrease of the target ES.
In most cases, the error rate is smaller than the target ES, reflecting that we finish the estimation when ES becomes smaller than the target ES.
When the target ES becomes smaller than $10^{-4}$, the fluctuation of the error rate in this evaluation becomes larger because the number of failure events becomes small.

For the further investigation of error rates, we calculate the noise ratio dependence of the evaluated error rate of the Bayesian and the averaging approach as shown in Fig~\ref{error}(b).
In the evaluation process, we set the target ES for the estimation as $10^{-3}$ and prepare datasets of $6.25 \times 10^{7}$ points for accurate evaluation.
The evaluated error rate is smaller than the target ES of $10^{-3}$.
By considering these results, we can utilize ES in the state estimation, realizing a specific error rate (the error rate is smaller than ES).
Especially, the error rate of the Bayesian approach is small at $\frac{\sigma_{0}}{\sigma_{1}}>1$.
Note that the behavior for state 1 can be described by exchanging the $\sigma_0$ and $\sigma_1$ in these results.

The error rates of the threshold judgment becomes
\begin{equation}
\begin{split}
{\rm ER}_{0,\rm thr} &= P(r\neq 0|q=0)_{\rm thr} \\
&= \frac{P_{0}\left[1-\frac{1}{2}\mathrm{erfc}(x_{0}) \right]}{\frac{P_{0}}{2}\mathrm{erfc}(x_{0})+\frac{P_{1}}{2}\mathrm{erfc}(x_{1})}
\label{eq:error rate0}
\end{split}
\end{equation}
\begin{equation}
\begin{split}
{\rm ER}_{1,\rm thr}&=P(r\neq 1|q=1)_{\rm thr} \\
&= \frac{\frac{P_{1}}{2}\mathrm{erfc}(x_{1})}{P_{0}\left[1-\frac{1}{2}\mathrm{erfc}(x_{0})\right]+P_{1}\left[1-\frac{1}{2}\mathrm{erfc}(x_{1})\right]}
\label{eq:error rate1}
\end{split}
\end{equation}
where, $x_q=\frac{t-V_{\mathrm{ RF}q}}{\sqrt{2}\sigma_q}$, $t$ is the threshold, $P_q$ the probability of obtaining $q$, and erfc the complementary error function.
Several ways to determine the optimal value of $t$ for each purpose are already known~\cite{Otsu, kittler1986minimum}, and we use the value of $t$ (which is a function of $P_{0,1}$) by calculating the total error rate ${\rm ER}_{0,\rm thr}+{\rm ER}_{1,\rm thr}$ should be minimum.
This error rate is plotted as green traces in Fig. ~\ref{Simulation}.
Because we already know the error rate of the Bayesian and the averaging approach is smaller than that of ES, we can compare the performance.
The proposed sequential estimation protocol is expected to be superior to the protocol based solely on histogram analysis, such as threshold judgment. 
This is because while threshold judgment always requires consistent performance regardless of the incoming values, our approach utilizes both time-dependent and histogram information.

\subsection{Real-time state estimation}
\begin{figure*}
\begin{center}
  \includegraphics{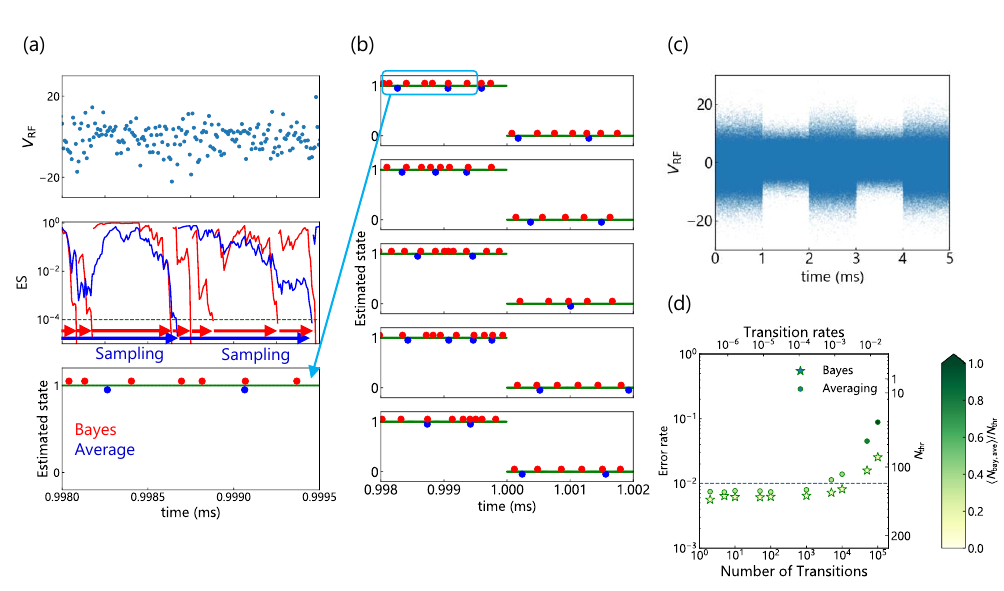}
  \caption{(a) Estimation protocol. By using the data points in the analysis, the error rates of the state estimation decrease.
  When the error becomes smaller than the threshold, we finish the estimation and create a new point in the estimated state, and then start the next state estimation.
  (b) Estimated charge state (0 or 1) as a function of time around the state transition by using averaging (blue) and Bayesian approach (red). Bayesian approach can extract more reliable data points. The five graphs show the results with five different data sets.
  (c) Test data for 5 ms with charge transitions every 1~ms. 
  (d) A number of transitions dependence of the error rate of the averaging approach and the Bayesian approach. A color bar indicates the ratio of the required number of data points and that in the threshold judgment satisfying the same error rate. The transition rate is the ratio of the number of transitions and total data points. 
  }
  \label{Realtime}
\end{center}
\end{figure*}

Next, we apply the Bayesian approach to extract the real-time change of the state by simulation.
In this estimation, we again request that the ES should be smaller than $10^{-4}$.
The condition of the simulation corresponds to $\sigma_0/\sigma_1=0.6$ and $\rm{SNR} = 0.4$.
The estimation protocol is shown in Figure~\ref{Realtime}(a). 
With increasing the number of data points in the analysis, the ES of the state estimation in both approaches decreases.
When the ES becomes smaller than $10^{-4}$, we finish the estimation and create a new point in the estimated state, and then start the next state estimation.
When the ES drops rapidly, for instance, under unbalanced noise conditions, the stopped value is significantly smaller than the requested ES.
It leads to a lower error rate compared with the requested ES, as shown in Fig~\ref{error}.
We prepare test data for 5 ms with every 8 ns with charge transitions in every 1~ms as shown in Figure~\ref{Realtime}(c).
Figure~\ref{Realtime}(b) shows the result of the estimation as a function of time by averaging approach (blue) and Bayesian approach (red). 
The five graphs show the results with five different data sets.
We can extract more data points in the Bayesian approach, reflecting the better estimation performance shown in Fig.~\ref{Simulation}(c) even near the state transitions.
The extracted points are not equally separated because we can calculate the probability iteratively and finish the estimation when the required ES is satisfied; estimation finishes quickly when we get good data points distributed close to $V_{\rm RF0}$ or $V_{\rm RF1}$.

We also evaluate the error rate of the Bayesian and averaging approaches with charge transitions by preparing data like Fig.~\ref{Realtime}(c).
Figure ~\ref{Realtime}(d) shows the number of transitions dependence of the error rate with the target ES of $10^{-2}$.
The error rate of both methods increases and goes above the target ES with the increase of the transition events.
But the typical required number of data points for the estimation $\langle N_{\rm bay} \rangle$ is always smaller than the number of points in threshold judgment $N_{\rm thr}$ satisfying the same error rate.
On the other hand, the typical required number of data points in the averaging approach $\langle N_{\rm ave} \rangle$ increases with increasing the number of transitions, and approaches $N_{\rm thr}$.
In particular, the Bayesian approach shows better performance than the averaging approach with many transitions reflecting the better estimation performance shown in Fig.~\ref{Realtime}(b) even near the state transitions.

\section{Discussion}
\subsection{Theoretical model of error score by Bayesian approach}
In order to explain why the Bayesian approach shows better performance in the unbalanced noise condition, we analyze the equation of the ES.
The ES in Bayesian approach $\rm ER_{\rm 0,bay}$ is calculated from Eq~\ref{eq:Bayes} as
\begin{equation}
\rm ES_{\rm 0,bay} = 1-P(0|\{ V_{{\rm RF\_} 0}, \ldots,V_{{\rm RF\_} n}\}) = \frac{1}{1+\alpha_0\frac{P_{\rm N}(0)}{P_{\rm N}(1)}}
\label{eq:Bayes2}
\end{equation}
$\rm ES_{\rm 0,bay}$ is the estimation accuracy corresponding to the probability of estimating state as ``1'' with input the state ``0''.
$\alpha_0$ is the acceleration term described by

\begin{equation}
\alpha_0 = \left(\frac{\sigma_1}{\sigma_0}\right)^{N-1}\mathrm{exp}\left[A_{\rm N}N\left(\frac{1}{{\sigma_1}^2} - \frac{1}{{\sigma_0}^2}\right) \right]
\label{eq:alpha}
\end{equation}

\begin{equation}A_{\rm N} = \frac{1}{2}\left[\frac{1}{N}\sum_{i=0}^N V_{\rm RFi}^2 - \left(\frac{1}{N}\sum_{i=0}^N V_{\rm RFi}\right)^2\right]
\label{eq:An}
\end{equation}
In the case of $\alpha_0 = 1$, $\rm ES_{\rm0, bay}$ is equal to $\rm ES_{\rm 0,ave}$ (Eq.~\ref{eq:Ave}).
When $\alpha_0 > 1$, $\rm ES_{\rm 0,bay}$ becomes smaller than $\rm ES_{\rm 0,ave}$.
From Eq.~\ref{eq:alpha}, the unbalanced noise situation leads to the large $\alpha_0$ with large $N$.
Because $\alpha_0$ increases with the increase of $N$, the Bayesian approach shows lower ES (see Appendix B). 
Furthermore, the $\alpha_0$ has two exponential terms with noise ratio and Euler's constant $e$ as the base.
When $\sigma_1/\sigma_0<1$,  the exponential term is dominant in $\alpha_0$.
On the other hand, the noise ratio term is dominant when $\sigma_1/\sigma_0>1$.
This produces the observed asymmetry in Fig.~\ref{Simulation}(d).

\section{Conclusion}
In conclusion, we propose the sequential charge state estimation method based on Bayes' theorem by simulations.
The Bayesian approach has a smaller error score compared to the averaging approach under the unbalanced noise condition between the empty and occupied states.
By reflecting better performance in the state estimation, the Bayesian approach can extract more reliable data points in the real-time state estimation and detect the faster change of the state.
Furthermore, we investigate the error rate of the Bayesian approach and find out that it has better performance compared with the averaging approach and threshold judgment. 
The proposed method is expected to be compatible with FPGAs for accelerated real-time processing, thereby enhancing single-electron detection crucial for qubit readout.

\section{Acknowledgements}
The authors thank J. Yoneda, S. Nagayasu, T. Nakajima, K. Takeda, A. Noiri, S. Tarucha, and RIEC Laboratory for Nanoelectronics and Spintronics 
for fruitful discussions and technical support.
Part of this work is supported by 
MEXT Leading Initiative for Excellent Young Researchers, 
Grants-in-Aid for Scientific Research (21K18592, 23H01789, 23H04490, 23KJ0200), 
Iketani Science and Technology Foundation Research Grant,
and FRiD Tohoku University.
Y. M. acknowledges WISE Program of AIE for financial support.

\section{Appendix A: Results in Cases with Frequency Dependent Noise}
We assume white noise in the main text. 
Here, we discuss results in cases with frequency-dependent noise.
We show that our methods can be effective even in these cases.
The frequency dependence of the noise reflects the frequency response of the resonator and the flicker noise due to device noise~\cite{shinozaki2021gate}.
\begin{figure*}[h]
\begin{center}
  \includegraphics[width=16.5cm]{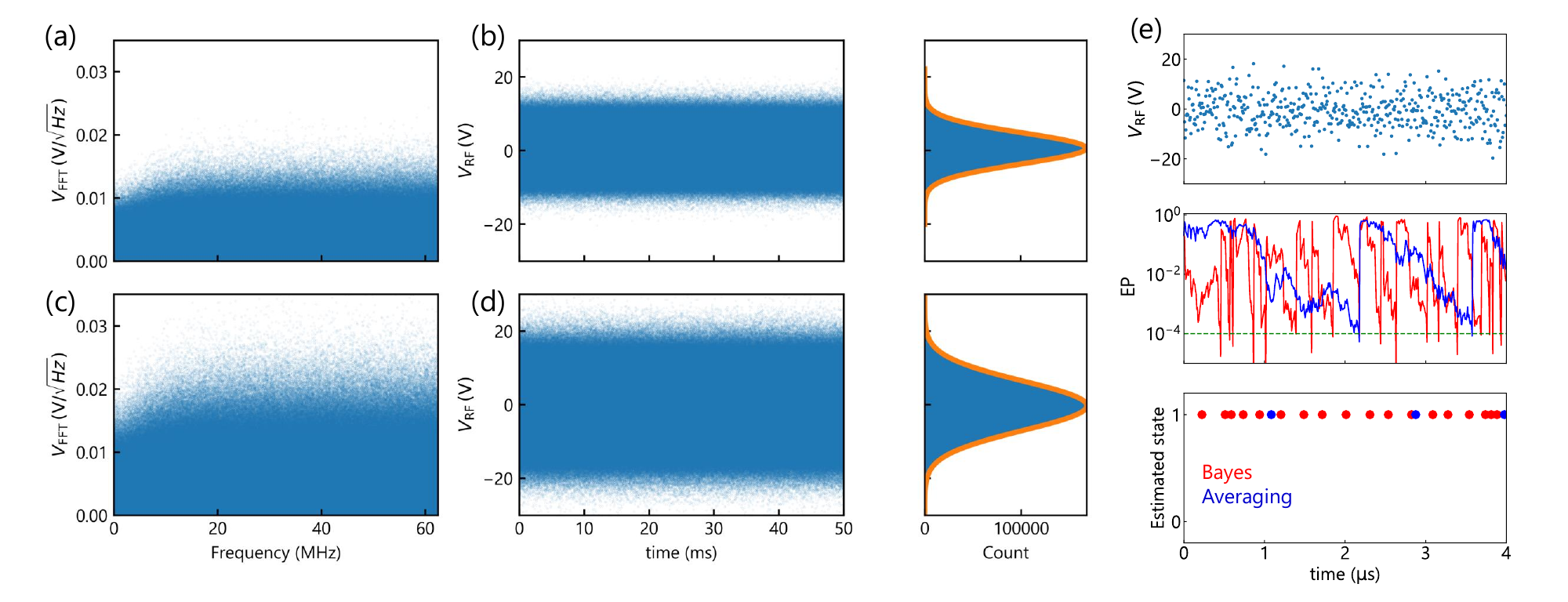}
  \caption{(a), (c) The noise spectrum considering the property of the tank circuit in RF reflectometry in 0 and 1 states. (b), (d) The simulated real-time data made by inverse-fast Fourier transform of the noise spectrum and its histogram. (e) The charge state estimation using the data in the case of state "1".
  }
  \label{noiseamp}
\end{center}
\end{figure*}

\begin{figure*}
\begin{center}
  \includegraphics[width=16.5cm]{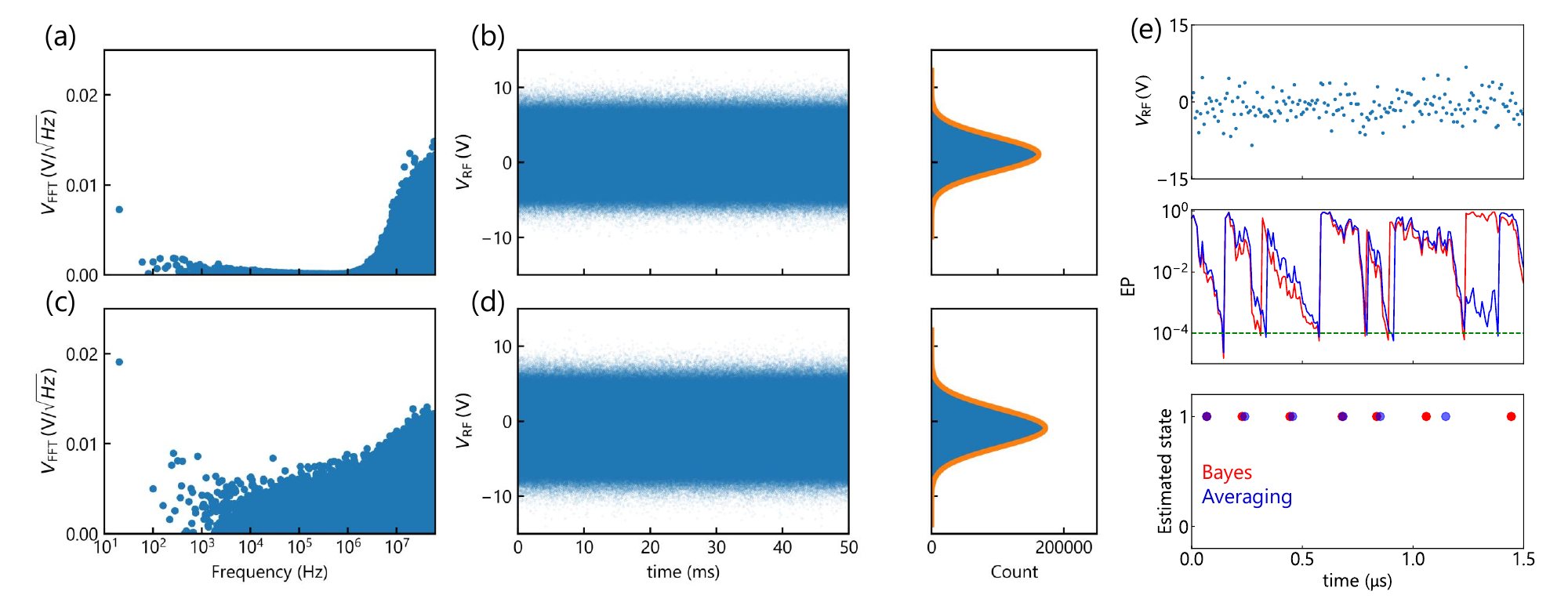}
  \caption{(a), (c) The noise spectrum considering the property of the resonator and the flicker noise are modulated by the charge state in 0 and 1 states. (b), (d) The simulated real-time data made by inverse-fast Fourier transform of the noise spectrum and its histogram. (e) The charge state estimation using the data in the case of state "1".
  }
  \label{noiseres}
\end{center}
\end{figure*}

\begin{figure}
\begin{center}
  \includegraphics{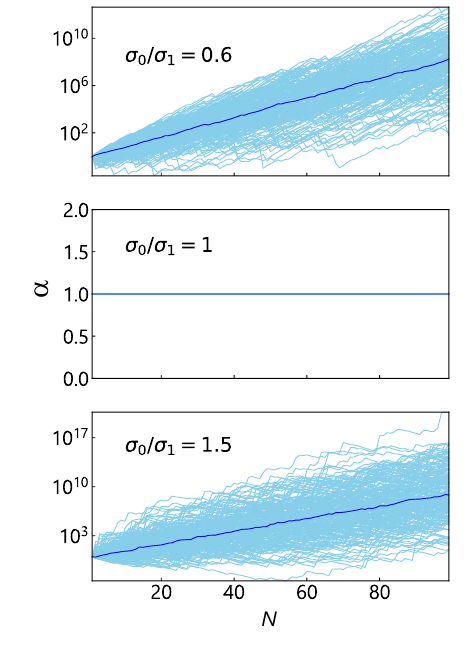}
  \caption{$\alpha$ as a function of the data points used for the charge state estimation $N$ in the case of SNR = 0.33 and $\sigma_{0}/\sigma_{1}=0.6$ (a), 1 (b) and 1.5 (c).
  }
  \label{alpha}
\end{center}
\end{figure}

At first, we assume that noise has different intensities in 0 and 1 states with the same frequency-dependence as shown in Figs~\ref{noiseamp} (a) and (c).
The frequency dependence comes from the property of the tank circuit in RF reflectometry.
We assume that the linewidth of the Lorentz function for the reflection coefficient and sampling rate are 10 MHz and 125 MHz, respectively.
The parameters are set as $\sigma_{0}/\sigma_{1}\approx 0.6$ and $\rm{SNR}=0.16$. 
We can extract around 9 times more reliable data points in the Bayesian approach (Fig~\ref{noiseamp} (e)).
This performance is almost identical to the case with the white noise.

Next, we consider the noise spectrum when the properties of the resonator and the flicker noise are modulated by the charge state (Fig~\ref{noiseres}).
This situation can be achieved using a highly sensitive charge sensor~\cite{eenink2019tunable, johmen2023radio} in conjunction with a low-noise amplifier~\cite{yamamoto2008flux}.
The difference in the frequency dependence also results in unbalanced noise corresponding to $\sigma_{0}/\sigma_{1}\approx 0.9$ and $\rm{SNR}=0.8$.
When we request the target ES of $10^{-3}$, $\langle N_{\rm bay} \rangle=19.5$ and $\langle N_{\rm ave} \rangle=19.9$ and these are 5 times faster than $N_{\rm thr}=95$ in the threshold judgement.
Note that the actual error rates of the Bayesian and the averaging approaches are evaluated to be $1.9\times 10^{-5}$ and $2.2\times 10^{-5}$, respectively. 
Even though the discrepancy of the actual error rates from the target value becomes worse, the charge-state estimation rate is still 5 times larger even when this level of frequency-dependent noise is assumed.
For further improvement of the estimation scheme, direct treatment of the frequency dependence in the calculation of the error probability will be a possible approach.

\section{Appendix B: $\alpha$ in Bayesian approach}

$\alpha$ makes the difference between the Bayesian and the averaging approaches.
Here, we discuss the details of $\alpha$.
Figure~\ref{alpha} shows $\alpha$ as a function of $N$ with SNR = 0.4 and $\sigma_{0}/\sigma_{1}=0.6$ (a), 1 (b) and 1.5 (c).
The light blue traces show the results with 200 different datasets.
The blue trace shows the median of the datasets.
If $\sigma_{0} \neq \sigma_{1}$, $\alpha$ increases with the increase of $N$.
Then, the error probability in the Bayesian approach becomes smaller than that in the averaging approach.
When $\sigma_{0} = \sigma_{1}$, $\alpha$ becomes 1, and both approaches show the same result.

\bibliography{reference}

\end{document}